\documentclass[preprints,article,accept,moreauthors,pdftex,10pt,a4paper]{Definitions/mdpi} 

\firstpage{1} 
\pubvolume{xx}
\issuenum{1}
\articlenumber{1}
\pubyear{2020}
\copyrightyear{2020}
\history{4 July 2020}

\pdfoutput=1

\usepackage{amssymb}
\usepackage[english]{babel}
\usepackage[utf8]{inputenc}
\usepackage{verbatim}
\usepackage{epsf}
\usepackage{epsfig}
\usepackage{nicefrac}

\Title{Parallel Lives: A~Local-Realistic Interpretation of ``Nonlocal'' Boxes\,%
\thanks{\,All the key ideas in this paper originate from the second author and are
based on his original doctoral work~\cite{PRRthesis} performed
under supervision of the first author, who participated in numerous
discussions and writing the paper.
The order of authors is alphabetic according to custom in the first
author's research community.
Even though there is a journal version of this paper, published as
\textit{Entropy} \textbf{21}(1):87 (2019), the present arXiv
version should be considered as \emph{the} authoritative version
because it contains some minor improvements in style
(compared to the \textit{Entropy} mandatory style) and contents.
Furthermore, the \textit{Entropy} version fails to be explicit
about the role of each author.
} }

\Author{Gilles Brassard $^{1,2}$ and Paul Raymond-Robichaud $^{3}$}

\AuthorNames{Gilles Brassard and Paul Raymond-Robichaud}

\address{$^{1}$ \quad D\'epartement d'informatique et de recherche op\'erationnelle, Universit\'e de Montr\'eal;
brassard@iro.umontreal.ca.\\
$^{2}$ \quad Canadian~Institute for Advanced Research, Toronto, Canada.\\
$^{3}$ \quad ISI Foundation, Torino, Italy; work done while at (1) above;\\
$^{\phantom{3}}$ \quad paul.r.robichaud@gmail.com.}

\abstract{We carry out a thought experiment in an imaginary world.
Our world is both local and realistic, yet it violates a Bell inequality more than does quantum theory.
This serves to debunk the myth that equates local realism with local hidden variables in the simplest possible manner.
Along the way, we reinterpret the celebrated 1935 argument of Einstein, Podolsky and Rosen,
and come to the conclusion that they were right in their questioning the completeness of
the Copenhagen version of quantum theory, provided one believes in a local-realistic universe.
Throughout our journey, we strive to explain our views from first principles, without expecting mathematical sophistication nor specialized prior knowledge from the reader. }


\begin{document}


\begin{raggedleft}
``\textsf{What is proved by impossibility proofs is lack of imagination}'' --- John Bell, 1982~\cite{BellQuote}\\
``\textsf{Imagination is more important than knowledge}'' --- Albert Einstein, 1929~\cite{EinsteinQuote}\\
\end{raggedleft}

\section{Introduction}\label{sc:intro}

Quantum theory is often claimed to be nonlocal, or more precisely that it cannot satisfy simultaneously the principles of locality and realism. These principles can be informally stated as follows.

\begin{description}
\item[Principle of realism:] There is a real world whose state determines the outcome of all observations.
\item[Principle of locality:] No action taken at some point can have any effect at some remote point at a speed faster than light.
\end{description}

A formal definition of local realism is given in a companion paper~\cite{STRUCTURE};
here we strive to remain at the intuitive level and explain all our concepts, results and reasonings
without expecting mathematical sophistication nor specialized prior knowledge from the reader.

The belief that quantum theory is nonlocal stems from the correct fact proved by John Bell~\cite{Bell64} that it cannot be described by a \emph{local hidden variable theory}, as we shall explain later.
However, the claim of nonlocality for quantum theory is also based on the incorrect
equivocation of local hidden variable theories with local realism, leading to the following fallacious argument: 

\newpage

\begin{enumerate}
\item\label{item:A} Any local-realistic world must be described by local hidden variables.
\item Quantum theory cannot be described by local hidden variables.
\item \emph{Ergo}, quantum theory cannot be both local and realistic.
\end{enumerate}
The first statement is false, as we explain at length in this paper; the second is true; the third is a legitimate application of \emph{modus tollens},\footnote{According to \emph{modus tollens}, if $p$ implies $q$ but $q$ is false, then $p$ must be false as well.} but the argument is unsound since
it is based on a false premise.
As~such, our reasoning does not imply that quantum theory can be both local and realistic, but it establishes decisively that the usual reasoning against the local realism of quantum theory is fundamentally flawed.

In~a companion paper, the second author explicitly derives a full and complete local-realistic interpretation for finite-dimensional unitary quantum theory~\cite{QMlocal}, which had already been discovered by David Deutsch and Patrick Hayden~\cite{DH}.
See also Refs.~\cite{Rubin,Tipler}.
Going further, he shows in another companion paper~\cite{STRUCTURE} that the local realism of quantum theory is but a particular case of the following more general statement: \emph{Any} reversible-dynamics theory that does not allow instantaneous signalling \mbox{admits} a local-realistic interpretation. 

In order to invalidate 
statement~(\ref{item:A}) above, we exhibit 
an imaginary world that is both local and realistic, yet that cannot be described by local hidden variables.  Our world is based on the so-called \emph{nonlocal box},
also known as the \emph{PR~box}, introduced by Sandu Popescu and Daniel Rohrlich~\cite{PR}, which is already known to violate a Bell inequality
even more than quantum theory (more on this later), which indeed implies that it cannot be explained by local hidden variables (more on this later also). Nevertheless, we provide a full local-realistic explanation for our imaginary world. Even though this world is not the one in which we live, its mathematical consistency suffices to debunk the myth that equates local realism with local hidden variables. In~conclusion,
the correct implication of Bell's theorem is that quantum theory cannot be described by local hidden variables, \emph{not} that it is not local-realistic. \mbox{\emph{That's different!}}

Given that quantum theory has a local-realistic interpretation, why bother with nonlocal boxes, which only exist in a fantasy world?
The main virtue of the current paper, compared to Refs.~\mbox{\cite{DH,Rubin,Tipler,QMlocal,STRUCTURE}}, is to invalidate the fallacious, yet ubiquitous, argument sketched above in the simplest and easiest possible way, without needing to resort to sophisticated mathematics.
The benefit of working with nonlocal boxes, rather than dealing with all the intricacies of quantum theory, was best said by Jeffrey Bub in his book on \emph{Quantum Mechanics for Primates}:
``The~conceptual puzzles of quantum correlations arise without the distractions of the mathematical formalism of quantum mechanics, and you can see what is at stake---where the clash lies with the usual presuppositions about the physical world''~\cite{Bub}.

The current paper is an expansion of an informal self-contained 2012 poster~\cite{poster} reproduced in the Appendix with small corrections,
which explains our key ideas in the style of a graphic novel,
as well as of a brief \mbox{account} in a subsequent paper~\cite{FreeWill}.
A~similar concept had already been formulated by
Mark A.~Rubin~\cite[p.~318]{Rubin} in the context of two distant observers measuring their shares of a Bell state
in the same basis, as well as Colin Bruce
in his popular-science book on \emph{Schr\"odinger's Rabbits}~\cite[pp.~130--132]{Rabbit}.
To~the best of our knowledge, the latter was the first local-realistic description of an imaginary world that cannot be described by local hidden variables.

After this introduction, we describe the Popescu-Rohrlich nonlocal boxes, perfect as well as imperfect, in Section~\ref{sc:NLB}. We~elaborate on no-signalling, local-realistic
and local hidden variable theories in Section~\ref{sc:nosig-locreal}, which we illustrate
with the Einstein-Podolsky-Rosen argu\-ment~\cite{EPR} and the nonlocal boxes.
Bell's Theorem is reviewed in Section~\ref{sc:Bell} in the context of nonlocal boxes, and we explain why they cannot be described by local hidden variables.
The~paper culminates with Section~\ref{sc:main}, in which we expound our theory of \emph{parallel lives} and how it allows us to show that ``nonlocal'' boxes are perfectly compatible with both locality and realism.
Having provided a solution to our conun\-drum, we~revisit Bell's Theorem and the Einstein-Podolsky-Rosen argument in Section~\ref{sc:revisit} in order to understand how they relate to our imaginary world.
There, we argue that our theory of parallel lives is an unavoidable consequence of postulating that the so-called nonlocal boxes are in fact local and realistic.
We~conclude with a discussion of our results in Section~\ref{sc:conc}.
Finally, we reproduce in the Appendix an updated version of our 2012 poster~\cite{poster}, which illustrates the main concepts.
Throughout our journey, we strive to illustrate how the arguments formulated in terms of nonlocal boxes and the more complex quantum theory are interlinked.

\section{The Imaginary World}\label{sc:NLB}
We~now proceed to describe how our imaginary world is perceived by its two inhab\-i\-tants, Alice and Bob.
We~postpone to Section~\ref{sc:main} a description of what is \emph{really} going on in that world.
The main ingredient that makes our world interesting is the presence of perfect nonlocal boxes,
a theoretical idea invented by Sandu Popescu and Daniel Rohrlich~\cite{PR}.

\subsection{The Nonlocal Box}
Nonlocal boxes always come in pairs: one box is given to Alice and the other to Bob.\,\footnote{Some people prefer to think of the nonlocal box as consisting of both boxes, so that the pair of boxes that we describe here constitutes a single nonlocal box. It's~a matter of taste.}
One can think of a nonlocal box as an ordinary-looking box with two buttons labelled $0$ and~$1$.
Whenever a button is pushed, the box instantaneously flashes either a red or green light,
with each outcome being equally likely.
This concept is illustrated in Figure~\ref{fig:PRbox}
and in the Appendix.
 
\begin{figure}[h!]
\newlength{\scale}
\setlength{\scale}{0.2pt}
\begin{center}
\includegraphics[width = 765\scale, height = 768\scale]{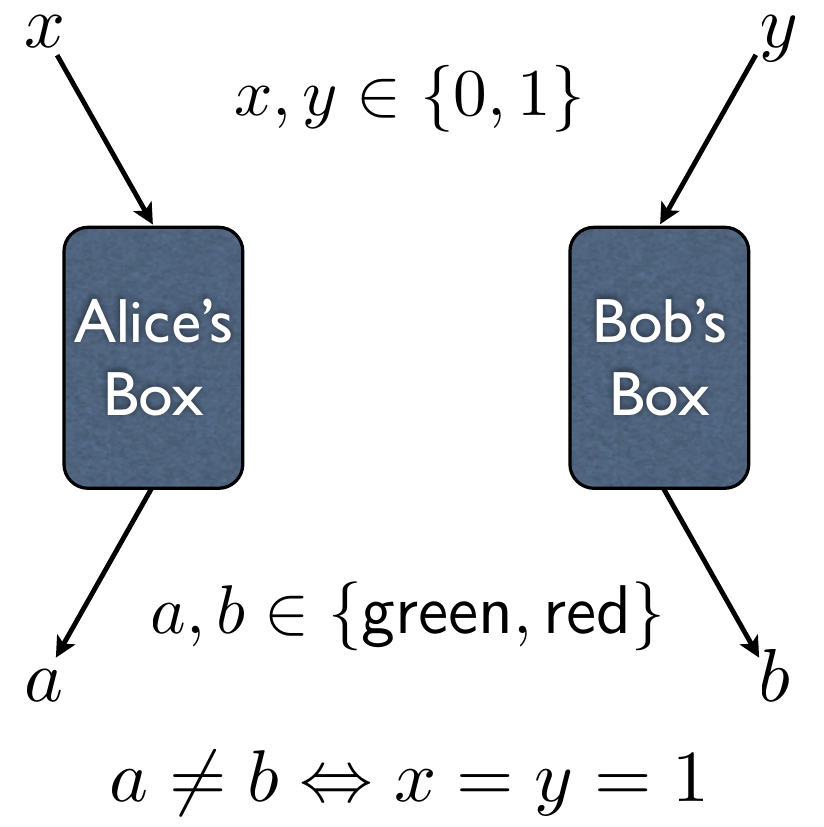}
\end{center}
\caption{Nonlocal boxes.}\label{fig:PRbox}
\end{figure}

If Alice and Bob meet to compare their results after they have pushed buttons, they will find that each
pair of boxes produced outputs that are correlated in the following way:
Whenever they had both pushed input button~$1$, their boxes flashed different colours,
but if at least one of them had pushed input button~$0$, their boxes flashed the same colour.
See~Table~\ref{tb:PRbox}.

\begin{table}[h!]
\caption{Behaviour of nonlocal boxes.}\label{tb:PRbox}
\[\begin{array}{ccc}
\text{Alice's Input} & \text{Bob's Input} & \text{Output colours}  \\ \hline
0 & 0 & \text{Identical} \\
0 & 1 & \text{Identical} \\
1 & 0 & \text{Identical} \\
1 & 1 & \text{Different} 
\end{array}
\]
\end{table}

For example, if Alice pushes $1$ and sees green, whereas Bob pushes~$0$, she will discover when she meets Bob that he has also seen green.  However, if Alice pushes $1$ and sees green (as~before), whereas Bob pushes~$1$ instead, she will discover when they meet that he has seen red.

A nonlocal box is designed for one-time use:  once a button has been pushed and a colour flashed, the box will forever flash that colour and is no longer responsive to new inputs.  However, Alice and Bob have an unlimited supply of such pairs of disposable nonlocal boxes.

\subsection{Testing the Boxes}\label{sc:test}
Our two inhabitants, Alice and Bob, would like to verify that their nonlocal boxes behave indeed according
to Table~\ref{tb:PRbox}. Here is how they proceed. 

\begin{enumerate}\itemsep02em
\item Alice and Bob travel far apart from each other with a large supply of numbered unused boxes, so that Alice's box number~$i$ is the one that is paired with Bob's box bearing the same number.
\item They flip independent unbiased coins labelled $0$ and $1$ and push the corresponding input buttons on their nonlocal boxes. 
For each box number, they record the randomly-chosen input and the observed resulting colour.
\mbox{Because} they are sufficiently far apart, the experiment can be performed with sufficient simultaneity that Alice's box cannot know the result of Bob's coin flip (hence the input to Bob's box) before it has to flash its own light, and vice versa.
\item After many trials, Alice and Bob come back together and verify that the boxes work perfectly: no matter how far they were from each other and how simultaneously the experiment is conducted, the correlations promised in Table~\ref{tb:PRbox} are realized for each and every pair of boxes.
\end{enumerate}

Note that neither Alice nor Bob can confirm that the promised correlations are estab\-lished until they meet, or at least send a signal to each other. In~other words, data collected locally at Alice's and at Bob's need to be brought together before any conclusion can be drawn.
This detail may seem insignificant at first, but it will turn out to be crucial in order to give a
local-realistic explanation for ``nonlocal'' boxes.

\subsection{Imperfect Nonlocal Boxes}
So far, we have talked about perfect nonlocal boxes, but we could consider nonlocal boxes that are sometimes allowed to give incorrectly correlated outputs. We~say that a pair of nonlocal boxes \emph{works with probability~$p$} if it behaves according to Table~\ref{tb:PRbox} with probability~$p$.
With complementary probability~\mbox{$1-p$}, the opposite correlation is obtained.

\subsubsection{Quantum Theory and Nonlocal Boxes}\label{sc:QT-vs-PRbox}
Although we shall concentrate on \emph{perfect} nonlocal boxes in this paper, quantum theory makes it possible to implement nonlocal boxes that work with probability
\[ \textstyle p_{\mathrm{quant}} = \cos^{2} \! \left( \frac{\pi}{8} \right) = \frac{2+\sqrt{2}}{4} \approx 85\%  \]
but no better according to Cirel'son's theorem~\cite{CIREL}.
It~follows that our imaginary world is distinct from the world in which we live since perfect nonlocal boxes cannot exist according to quantum theory.
 
For our purposes, the precise mathematics and physics that are needed to under\-stand how it is possible for quantum theory to implement nonlocal boxes that work with probability~$p_{\mathrm{quant}}$ do not matter.
Let us simply say that it is made possible by harnessing \emph{entanglement} in a clever way.
Entanglement, which is the most nonclassical of all quantum resources, is at the heart of
quantum information science.
It~was discovered by Einstein, Podolsky and Rosen in 1935 in Einstein's most cited paper~\cite{EPR},
although there is some evidence that Erwin Schr\"odinger had discovered it earlier.
It~is also because of entanglement that the quantum world in which we live is often thought
to be nonlocal.

\section{The Many Faces of Locality}\label{sc:nosig-locreal}

Recall that the Principle of locality claims that no action taken at some point can have any effect at some remote point at a speed faster than light. An~apparently weaker principle would allow such effects under condition that they cannot be observed at the remote point. This is the Principle of no-signalling, which we now explain.

\subsection{No-signalling}
It is important to realize that nonlocal boxes do not enable instantaneous communication between Alice and Bob. Indeed, no matter which button Alice pushes (or~if she does not push any button at~all),
Bob has an equal chance of seeing red or green flashing from his box whenever he pushes either of his buttons. Said other\-wise, any action taken by Alice has no effect whatsoever on the probabilities of events that Bob can observe.

It follows that our imaginary world shares an important property with the quantum world:
it obeys the principle of no-signalling.

\begin{description}
\item[Principle of no-signalling:] No action taken at some point can have any \emph{observable} effect at some remote point at a speed faster than light.
\end{description}

Among observable effects, we include anything that would affect the probability distribution of outputs from any device.  The principle of no-signalling implies in general that for any pair of devices shared by Alice and Bob (not only PR boxes), Bob's output distribution depends only on Bob's input, and not on Alice's input, provided they are sufficiently far from each other and Alice does not provide her input to her device too long before Bob's device must produce its output.

\subsection{Local Realism Implies No-signalling}
The principle of no-signalling follows from the principles of locality and realism:
any local-realistic world is automatically no-signalling, as shown by the following informal argument.

\begin{enumerate}
\item By the principle of locality, no action taken at point $A$ can have any effect on the state of the world at point $B$ faster than at the speed of light.
\item By the principle of realism, anything observable at point $B$ is a function of the state of the world at that point.
\item It~follows that no action at point $A$ can have an observable effect at point $B$ faster than at the speed of light.
\end{enumerate}

Here, we have relied on the tacit assumption that in a local-realistic world, what is
observable at some point is a function of the state of the world at that same point.
The above argument is fully formalized, with all hypotheses made explicit, in Ref.~\cite{STRUCTURE}.

\subsection{Local Hidden Variable Theories}

The most common type of local-realistic theories, which were studied in particular by John Bell~\cite{Bell64},
is based on local hidden variables (explained below). The misconception according to which all local-realistic theories have to be of that type has led to the widespread misguided belief that quantum theory cannot be local-realistic because it cannot be based on local hidden variables according to Bell's theorem.

In the idealized context of nonlocal boxes, a local hidden variable theory would consider arbitrarily sophisticated pairs of devices that are allowed to share randomness for the purpose of explaining the observed behaviour. The individual boxes would also be allowed independent sources of internal randomness.
The~initial shared randomness, along with the internal randomness and the inputs provided by Alice and Bob would be used to determine which colours to flash. \mbox{However}, what is \emph{not} allowed is for the output of one of Alice's boxes to depend on the input of Bob's corresponding box, or vice versa. This can be enforced by the principle of locality, provided both input buttons are pushed with sufficient simul\-ta\-neity to prevent a signal from one box to reach the other in time, even at the speed of light, to influence its outcome.

In any such theory (not just those pertaining to PR boxes), it is always possible to remove internal sources of randomness and replace them by parts of the initial source of shared randomness that would be used by one side only
(provided we allow an infinite amount of shared randomness).
However, the following section shows that, in the case of \emph{perfect} nonlocal boxes, internal randomness should \emph{never} be used to influence the behaviour of PR boxes.

\subsection{The Einstein-Podolsky-Rosen Argument}\label{sc:EPR}

Even though they were obviously not talking about Popescu-Rohrlich nonlocal boxes,
the original 1935 argument of Einstein, Podolsky and Rosen (EPR) applies \emph{\mbox{mutatis} mutandis} to prove
that, in the context of local hidden variable theories, the output of Bob's nonlocal box should be completely determined by the initial randomness shared between Alice's and Bob's boxes and by Bob's input (and vice versa, with Alice and Bob interchanged).

\begin{enumerate}
\item Suppose Alice pushes her input button first.\,\footnote{For simplicity, we ignore the fact that there would be no such thing as absolute time if we took account of relativity, so that the notion of who pushes the button first may be ill-defined; this has no impact on the current reasoning because it is well-defined whether the effect of a button push can reach the other side before the other button is pushed.}
\item When she pushes her button, this cannot have any instantaneous effect on Bob's box, by the principle of locality.
\item After seeing her output, Alice can know with certainty what colour Bob will see as a function of his input (even though she does not know which input he will choose). For example, if Alice had pushed $1$ and seen green, she knows that if Bob chooses to push $0$ he will also see green, whereas if he chooses to push $1$ he will see red.
\item Since it is possible for Alice to know with certainty what colour Bob will see when he pushes either button, and she can obtain this knowledge without  influencing his system, it must be that his colour was  \emph{predetermined} as a function of which button he would push.
This predetermination can only come from the initial source of shared randomness, and errors could occur if it were influenced by local randomness at Bob's.
\end{enumerate}

This argument was used in the original Einstein-Podolsky-Rosen \mbox{paper}~\cite{EPR}
to prove, under the implicit assumption of local hidden variables, that there are instances in quantum theory
in which both the position and the momentum of a particle must be simultaneously defined.
This clashed with the Copenhagen vision of quantum theory, according to which Heisenberg's uncertainty principle arises not merely from the fact that measuring one of those properties necessarily disturbs the other, but because they can never be fully defined simultaneously. 
The~conclusion of Einstein, Podolsky and Rosen was that (the~Copen\-hagen) quantum-mechanical descrip\-tion of physical reality can\emph{not} be considered complete.
After Niels Bohr's response~\cite{Bohr}, the physics community consensus was largely in his favour,
asserting that the EPR argument was unsound and that the Copenhagen interpretation is indeed complete.
In~a companion paper~\cite{QMlocal}, we prove that, under the metaphysical principle
of local realism, it is Einstein, Podolsky and Rosen who were correct after all in arguing that the usual formulation of quantum theory cannot be a complete descrip\-tion of physical reality, and furthermore we provide a solution to make it complete along lines similar to those already discovered by Deutsch and Hayden~\cite{DH}.

But let us come back to the imaginary world of nonlocal boxes\ldots

\subsection{Local Hidden Variable Theory for Nonlocal Boxes}
In a local hidden variable theory for nonlocal boxes, we have seen that all correlations should be explained by the initial shared randomness. Since each box implements a simple one-bit to one-colour-out-of-two function, it suffices to use only two bits of the randomness shared with its twin box to do~so. It~is natural to call those bits $A_0$ and $A_1$ for Alice, and $B_0$ and $B_1$ for Bob. If~we define function
\[ \mathsf{c} : \{ 0,1 \} \rightarrow \{ \mathsf{green}, \mathsf{red} \} \]
by $\mathsf{c}(0)=\mathsf{green}$ and $\mathsf{c}(1)=\mathsf{red}$,
then Alice's box would flash colour \mbox{$a=\mathsf{c}(A_x)$} when input button~$x$ is pushed by Alice, whereas
Bob's box would flash colour \mbox{$b=\mathsf{c}(B_y)$} when input button~$y$ is pushed by Bob.
See~Figure~\ref{fig:PRbox} again.

In order to fulfil the requirements of nonlocal boxes given in Table~\ref{tb:PRbox},
it is easy to verify that
the four local hidden variables must satisfy the condition
\begin{equation}\label{eq:PRbox}
 A_{x} \oplus B_{y} = x \cdot y
\end{equation}
for all \mbox{$x,y \in \{ 0,1 \}$} simultaneously,
where ``$\oplus$'' and ``$\cdot$'' denote the sum and the product \mbox{modulo~2}.
For~example, if Alice and Bob select \mbox{$x=0$} and \mbox{$y=1$}, respectively, their boxes must flash the same colour
\mbox{$a=\mathsf{c}(A_0)=\mathsf{c}(B_1)=b$}, according to Table~\ref{tb:PRbox}, and therefore the hidden variables $A_0$ and
$B_1$ must be equal since function $\mathsf{c}$ is one-to-one.
In~symbols, \mbox{$A_{0}  = B_{1}$},  which is equivalent to
\mbox{$A_{0} \oplus B_{1} = 0$}, which indeed is equal to \mbox{$x \cdot y$} in this case.                               
                                 
Is this possible?

\section{Bell's Theorem}\label{sc:Bell}
\begin{Theorem}[Bell's Theorem]\label{th:Bell} No local hidden variable theory can explain a nonlocal box that would work with a probability better than~$75\%$. 
In~particular, no local hidden variable theory can explain perfect nonlocal boxes.
\end{Theorem}
\begin{proof}
We have just seen that any local hidden variable theory that enables perfect nonlocal boxes would have to satisfy Equation~ (\ref{eq:PRbox}) for all \mbox{$x,y \in \{ 0,1 \}$}.
This gives rise to the following four explicit equations.
\begin{align*}
A_{0} \oplus B_{0} &= 0  \\
A_{0} \oplus B_{1} &= 0  \\
A_{1} \oplus B_{0} &= 0  \\
A_{1} \oplus B_{1} &= 1  
\end{align*}
If we sum modulo~2 the equations on both sides and rearrange the terms
using the associativity and commutativity of addition modulo~2, as well as the fact that any bit added modulo~2 to itself gives~$0$, we obtain:
\begin{align*}
 \left( A_{0} \oplus B_{0} \right) \oplus  \left( A_{0} \oplus B_{1} \right) \oplus \left( A_{1} \oplus B_{0} \right) \oplus \left( A_{1} \oplus B_{1} \right) &=   0 \oplus 0 \oplus 0 \oplus 1  \\
 \left( A_{0} \oplus A_{0} \right) \oplus \left( A_{1} \oplus A_{1} \right) \oplus \left( B_{0} \oplus B_{0} \right) \oplus \left( B_{1} \oplus B_{1} \right) &=  1  \\
 0 \oplus 0 \oplus 0 \oplus 0   &=   1 \\
 0  & =     1 ,
\end{align*}
which is a contradiction.  Therefore, it is not possible for all four equations to hold simultaneously. At~least one of the four possible choices of buttons pushed by Alice and Bob is bound to give incorrect results.   It~follows that any attempt at creating a nonlocal box that works with probability better than $\nicefrac{3}{4}= 75\%$ is doomed to fail in any theory based on local hidden variables.
\end{proof}

The reader can easily verify from the proof of Theorem~\ref{th:Bell} that any three of the four equations can be satisfied by a proper choice of local hidden variables. 
For~example, setting \mbox{$A_{0} = B_{0} = A_{1} = B_{1} = 0$} results in the first three equations being satisfied, but not the fourth. A~more interesting strategy would be for Alice's box to produce \mbox{$A_{x}=x$} and for Bob's box to produce \mbox{$B_{y}=1-y$}.
In~this case, the last three equations are satisfied but not the first.
For each equation, there is a simple strategy that satisfies the other three but not that one
(more than one such strategy in fact).
More interestingly, it is possible to create a pair of nonlocal boxes that works
with probability $75\%$ \emph{regardless of the chosen input}
if the boxes share three bits of randomness.
The first two bits determine which one of the four equations is jettisoned, thus defining an arbitrary pre-agreed strategy
that fulfils the other three. If~the third random bit is $1$, both boxes will in fact produce the complement of the output specified in their strategy (which has no effect on which equations are satisfied). The purpose of this third shared random bit is that a properly functioning pair of PR boxes should produce an unbiased random output on each side if we only consider marginal probabilities.

We say of any world in which nonlocal boxes exist that work with a probability better than $75\%$ that it \emph{violates a Bell inequality} in honour of John Bell, who established the first result along the lines of Theorem~\ref{th:Bell}, albeit not explicitly the one described here~\cite{Bell64}.

\subsection{Quantum Theory and Bell's Theorem}

The usual conclusion from Theorem~\ref{th:Bell} is that any world containing nonlocal boxes that work with a probability better than $75\%$ cannot be both local and real\-is\-tic. Since quantum theory enables boxes that work $\approx 85\%$ of the time, as we have seen in Section~\ref{sc:QT-vs-PRbox}, it seems inescapable that the quantum world cannot be local-realistic.

Similarly, it is tempting to assert that the more a Bell inequality is violated by a theory, the more nonlocal it~is.
In~particular, our imaginary world would be more nonlocal than the quantum world itself.
As~we shall now see---and this is the main point of this paper---all these conclusions are unsound because  local realism and local hidden variables should not be equated.

\section{A Local Realistic Solution---Parallel Lives}\label{sc:main}

Here is how the seemingly impossible can be accomplished.
Let~us assume for simplicity that Alice and Bob have a single pair of ``nonlocal'' boxes at their disposal,
which is sufficient to rule out local hidden variable explanations.
When Alice pushes a button on her box, she splits in two, together with her box. One Alice sees the red light flash on her box, whereas the other sees the green light flash. Both Alices are equally real.  However, they are now living \emph{parallel lives}: they will never be able to see each other or interact with each other. 
In~fact, neither Alice is aware of the existence of the other, unless they infer it by pure thought as the
only local-realistic explanation for what they will experience when they test their boxes according to Section~\ref{sc:test}.
From now on, any unsplit object (or person) touched by either Alice or her box splits and inherits the splitting power.
This does not have to be direct physical touching: a message sent by Alice has the same splitting effect on anything it reaches. Hence, Alice's splitting ripples through space, but no faster than at the speed of light. It~is crucial to understand that it is \emph{not} the entire universe that splits instantaneously when Alice pushes her button, as this would be a highly nonlocal effect.

The same thing happens to Bob when he pushes a button on his box: he~splits and neither copy is aware of the other Bob.  One copy sees a red light flash and the other sees a green light flash. If~both Alice and Bob push a button at about the same time, we have two independent Alices and two independent Bobs,
and for now the Alices and the Bobs are also independent of one another.

It is only when Alice and Bob interact that correlations are established.
Let us assume for the moment that both Alice and Bob
always push their buttons before interacting.
The magical rule is that an Alice is allowed to interact with a Bob if and only if they jointly satisfy the conditions of the nonlocal box set out in Table~\ref{tb:PRbox}.

For example, if Alice pushes button~$1$, she splits. Consider the Alice who sees green.
Her system can be imagined to carry the following rule: You are allowed to interact with Bob if either he had pushed button~$0$ on his box and seen green, or pushed button~$1$ and seen red.
Should this Alice ever come in presence of a Bob who had pushed button~$1$ and seen green,
she would simply not become aware of his presence and could walk right through him without either one of them noticing anything.
Of course, the other Alice, the one who had seen red after pushing button~$1$, would be free to shake hands with that Bob.

If Bob had pushed button~$0$ and seen green, his system can likewise be imagined to carry the following rule: You are allowed to interact with Alice if and only if she sees green, regardless of which button she had pushed. It~is easy to generalize this idea to all cases covered by Table~\ref{tb:PRbox} because there will  always be one green Alice and one red Alice, one green Bob and one red Bob, and whenever green Alice is allowed to interact with one Bob, red Alice is allowed to interact with the other Bob.
From their perspective, each Alice and each Bob will observe correlations that seem to ``emerge from outside space-time''~\cite{GisinFreeWill}. However, this inter\-pre\-ta\-tion is but an illusion due to their intrinsic inability to perceive some of the actors in the world in which they live.

Our imaginary world is fully local because Alice's state is allowed to depend only on her own input and output at the moment she pushes a button.
It~is true that the mysterious correlations given in Table~\ref{tb:PRbox} would be impossible for any local hidden variable theory. However, Alice and Bob cannot \emph{experience} those correlations before they actually meet (or~at least before they share their data),
\emph{and these encounters cannot take place faster than at the speed of light.}
When they meet, the correlations they experience are simply due to the
matching rule that determines which Alices are allowed to interact with which Bobs,
and \emph{not} to a magical (because nonlocal) spukhafte Fernwirkung (``spooky action at a distance''), which was so abhorrent to Einstein, and rightly~so.

What if Alice pushes her button, but Bob does not?
In~the discussion above, we assumed for simplicity that both Alice and Bob had pushed buttons on their boxes before interacting. A~full story should include various other scenarios. It~could be that Alice pushes a button on her box and travels to interact with a Bob who had not yet touched his box. Or~it could be that after pushing a button on her box, only the Alice who had seen green travels to interact with Bob, whereas the Alice who had seen red stays where she~is.

For instance, consider the case in which
Alice had pushed button~$1$ on her box, split, and only the Alice who had seen green travels to meet unsplit Bob. At~the moment they meet, Bob and his box automatically split.
One Bob now owns a box programmed as follows: ``if~button~$0$ is pushed, flash green, but if button~$1$ is pushed, flash~red''; the other Bob owns a box containing the complementary program, with ``green'' and ``red'' interchanged.
As~for our travelling Alice, she will see the first one of those Bobs and be completely oblivious of the other, who will not even be aware that an Alice had just made the trip to meet him.

It would be tedious, albeit elementary, to go through an exhaustive list of all possible scenarios.
We~challenge the interested reader to figure out how to make our imaginary world behave according to Table~\ref{tb:PRbox} in all cases.
But~rather than get bored at this exercise, why not enjoy the Appendix,
which illustrates the concept of parallel lives in the form of a graphic novel~\cite{poster}?

\subsection{Quantum Theory, Parallel Lives and Many Worlds}

We coined the term ``parallel lives'' for the idea that a system is allowed to be in a superposition of several states, but that all splittings occur locally.
This was directly inspired by the many-worlds interpretation of quantum theory,
whose pioneer was Hugh Everett~\cite{Everett} more than six decades ago.
However, our parallel lives theory is fundamentally distinct from the highly nonlocal---and very
popular~\cite[pp.~119--121]{Random}---na\"{\i}ve version of its many-worlds counterpart according to which the entire universe would
split whenever Alice pushes a button on her box (or~makes a measurement that has more than one possible outcome according to standard quantum theory).
Later, Deutsch and Hayden provided the first explicitly local formulation of quantum theory, including a very lucid explanation of why Bell's theorem is irrelevant~\cite{DH}. Even though they did not use the term ``parallel lives'', their approach was akin to ours. In~their solution, the evolution of the quantum world is fully local, and individual systems, including observers, are implicitly allowed to be in superposition.
In~a companion paper~\cite{QMlocal}, we offer our own local formalism for quantum theory along the lines of this paper, complete with full proofs of our assertions.

\section{Revisiting Bell's Theorem and the Einstein-Podolsky-Rosen Argument}\label{sc:revisit}

Having provided a solution to our conundrum with the explicit construction of a local-realistic imaginary world in which perfect Popescu-Rohrlich ``nonlocal'' boxes are possible, we~revisit the Einstein-Podolsky-Rosen argument in order to understand how it relates to our imaginary world. This leads us to conclude that our theory of parallel lives is an unavoidable consequence of postulating that those boxes are compatible with local realism.

\subsection{Parallel Lives versus Hidden Variable Theories}
To understand the main difference between parallel lives and local hidden variable theories, consider again
the scenario according to which Alice had pushed button~$1$ and her box flashed a green colour.
According to local hidden variable theories, she would know with certainty what colour Bob will see as a function of his choice of input: he~will also see green if he pushes button~$0$, but he will see red if he pushes button~$1$.
This was at the heart of the Einstein-Podolsky-Rosen argument of Section~\ref{sc:EPR} to the effect that the colours flashed by Bob's box had to be predetermined as a function of which button he would push since Alice could know this information without interacting with Bob's box. To~quote the original argument,
``If,~without in any way disturbing a system, we can predict with certainty the value of a physical quantity, then there exists an element of physical reality corresponding to this physical quantity''~\cite{EPR}.
The~``element of physical reality'' in question is what we now call local hidden variables and the
``physical quantity'' is the mapping between input buttons and output colours.

The parallel-lives interpretation is fundamentally different.
Whenever Alice pushes a button on her box, she cannot infer anything about Bob's box.
Instead, she can predict how her various lives will interact with Bob's in the future,
in case they meet.
Consider for example a situation in which both Alice and Bob push their input buttons, whose immediate effect is the creation of two Alices and two Bobs. Let~us call them Green-Alice, Red-Alice, Green-Bob and Red-Bob,
depending on which colour they have seen.
If~the original Alice had pushed her button~$1$, Green-Alice may now infer that she will interact with Red-Bob if he had also pushed his button~$1$, whereas she will interact with Green-Bob if he had pushed his button~$0$. The~opposite statement is true of Red-Alice. As~we can see, this is a purely local process since this instantaneous knowledge of both Alices has no influence on whatever the faraway Bobs may observe, which is actually both colours!

\subsection{How an Apparent Contradiction Leads to Parallel Lives}

Consider the following argument concerning nonlocal boxes, and pretend that you have never heard
of parallel lives (nor of many worlds), yet you believe in locality.

\begin{enumerate}
\item\label{it:one} Let us say that Alice pushes button~$1$ on her box. Without loss of generality, say that her box flashes the green colour.
\item\label{it:two} Now, we know that Bob will see green if he pushes his button~$0$, whereas he will see red if he pushes his button~$1$, according to Table~\ref{tb:PRbox}.  By~the principle of locality, this conclusion holds regardless of Alice's previous action since she was too far for her choice of button to influence Bob's box.
\item\label{it:three} What would have happened had Alice pushed her button~$0$ instead at step~\ref{it:one}?
She must see the same colour as Bob, regardless of Bob's choice of button, since her pushing button~$0$ precludes the possibility that both Alice and Bob will press their button~$1$, which is the only case yielding different colours, again according to Table~\ref{tb:PRbox}.
\item\label{it:four} Statements (\ref{it:two}) and (\ref{it:three}) imply together that when Alice pushes her button~$0$, she must see both red and green!
\end{enumerate}

Despite appearances, statement~\ref{it:four} is not a contradiction, and indeed it can be resolved.
Both results seen by Alice must be equally real by logical necessity.
The~only way for her to see both colours, \emph{yet be convinced she saw only one}, is that there are in fact two Alices unaware of each other. In~other words, the postulated locality of Popescu-Rohrlich ``nonlocal'' boxes
\emph{forces us} into 
a parallel-lives theory, which,
far from being a postulate, is in fact an ineluctability.

If both Alices are indeed mathematically necessary to describe a local-realistic world, then both Alices are real in that world, inasmuch as any mathematical quantity that is necessary to describe reality corresponds to something that is real. Here, we accepted as a philosophical axiom the claim that whenever a mathematical quantity is necessary to describe reality, that quantity  corresponds to something that is real, and is not a mere artifact of the theory.

The same conclusion applies whenever any theory is shown to be inconsistent with all possible local hidden variable theories. Indeed, such theories carry the rarely-mentioned assumption that once concluded, any experiment has a single outcome. Other outcomes that could have been possible simply did not occur.
The~inescapable resolution of any such inconsistency is to accept the conclusion that all possible outcomes occur within parallel lives of the experimenter.

\section{Conclusions}\label{sc:conc} 
We~have exhibited a local-realistic imaginary world that violates a Bell inequality.
For~this purpose, we introduced the concept of \emph{parallel lives}, but argued subsequently that 
this was an unavoidable consequence of postulating that the so-called nonlocal boxes are in fact local and realistic.
The main virtue of our work is to demonstrate in an exceedingly simple way that local reality can produce correlations that are impossible in any theory based on local hidden variables.
In~particular, it is fallacious to conclude that quantum theory is nonlocal simply because it violates Bell's inequality.

In quantum theory, ideas analogous to ours can be traced back at least to Everett~\cite{Everett}.
They were developed further by Deutsch and Hayden~\cite{DH}, and subsequently by
Rubin~\cite{Rubin} and Tipler~\cite{Tipler}. Furthermore, Bruce \mbox{\cite[pp.~130--132]{Rabbit}} gave the first local-realistic explanation for a theory that is neither quantum nor classical.
In~companion papers, we have proven that unitary quantum mechanics is local-realistic~\cite{QMlocal} (which had already been shown in Ref.~\cite{DH}) and,
more generally, that this is true for \emph{any} reversible-dynamics no-signalling operational theory~\cite{STRUCTURE}.
The latter paper provides a host of suggestions in its final
section for a reader eager to pursue this line of work in yet unexplored directions.

Throughout our journey, we have revisited several times the
Einstein-Podolsky-Rosen argument and have come to the conclusion that they were right in questioning the completeness of Bohr's Copenhagen quantum theory.
Perhaps Einstein was correct in his belief of a local-realistic universe after all and in wishing for quantum theory to be completed?
Perhaps we live parallel lives\ldots


\vspace{6pt} 

\funding{The work of G.B. is supported in part by the Canadian Institute for Advanced Research, the Canada Research Chair program,
Canada's Natural Sciences and Engineering Research Council (\textsc{nserc})
and Qu\'ebec's Institut transdisciplinaire d'information quantique.
The work of P.R.-R. was supported in part by \textsc{nserc}
and the Fonds de recherche du Qu\'ebec -- Nature et technologies.}

\acknowledgments{We acknowledge stimulating discussions with Charles Alexandre B\'edard, Charles Bennett, Jeff Bub, Giulio Chiribella, David Deutsch, St\'ephane Durand, Marcin Paw{\l}owski, Sandu Popescu, Renato Renner, Alain Tapp and Stefan Wolf,
as well as careful proofreadings by Yuval Elias.
Furthermore, we acknowledge the artwork of Louis Fernet-Leclair, who drew the poster reproduced in the Appendix according to our specifications.
We~are also grateful to Lev Vaidman, whose special issue on Quantum Nonlocality for \textit{Entropy}
prompted us to finally commit to paper the second author's old 2012 idea~\cite{poster,FreeWill}.
In~his Editorial for the corresponding \textit{Entropy} special issue~\cite{Vaidman}, Vaidman wrote:
``Very subjectively---I find the most interesting contribution to be the work by Brassard and Raymond-Robichaud''.
We~also acknowledge some insightful suggestions from the anonymous \textit{Entropy} referees.
Finally, G.B.~is grateful to Christopher Fuchs for having dragged him \emph{up} the slope of quantum foundations years ago, despite their subsequent divergent paths.}


\abbreviations{The following abbreviations are used in this manuscript:\\
\noindent 
\begin{tabular}{@{}ll}
EPR & Einstein-Podolsky-Rosen\\
PR & Popescu-Rohrlich
\end{tabular}}

\reftitle{References}
\externalbibliography{yes}
\bibliography{references}

\appendixtitles{yes}

\appendixsections{one}

\pagebreak

\appendix

\section{Poster on Parallel Lives}

We reproduce below the 2020 improved version of the poster realized by Louis Fernet-Leclair in 2012
according to our specifications~\cite{poster}. Higher-resolution versions are available at
\url{http://www.iro.umontreal.ca/~brassard/ParallelLives}.

\nopagebreak

\bigskip

\noindent
\newlength{\scaleposter}
\setlength{\scaleposter}{0.175pt}
\hspace*{-25pt}
\includegraphics[width = 2774\scaleposter, height = 3500\scaleposter]{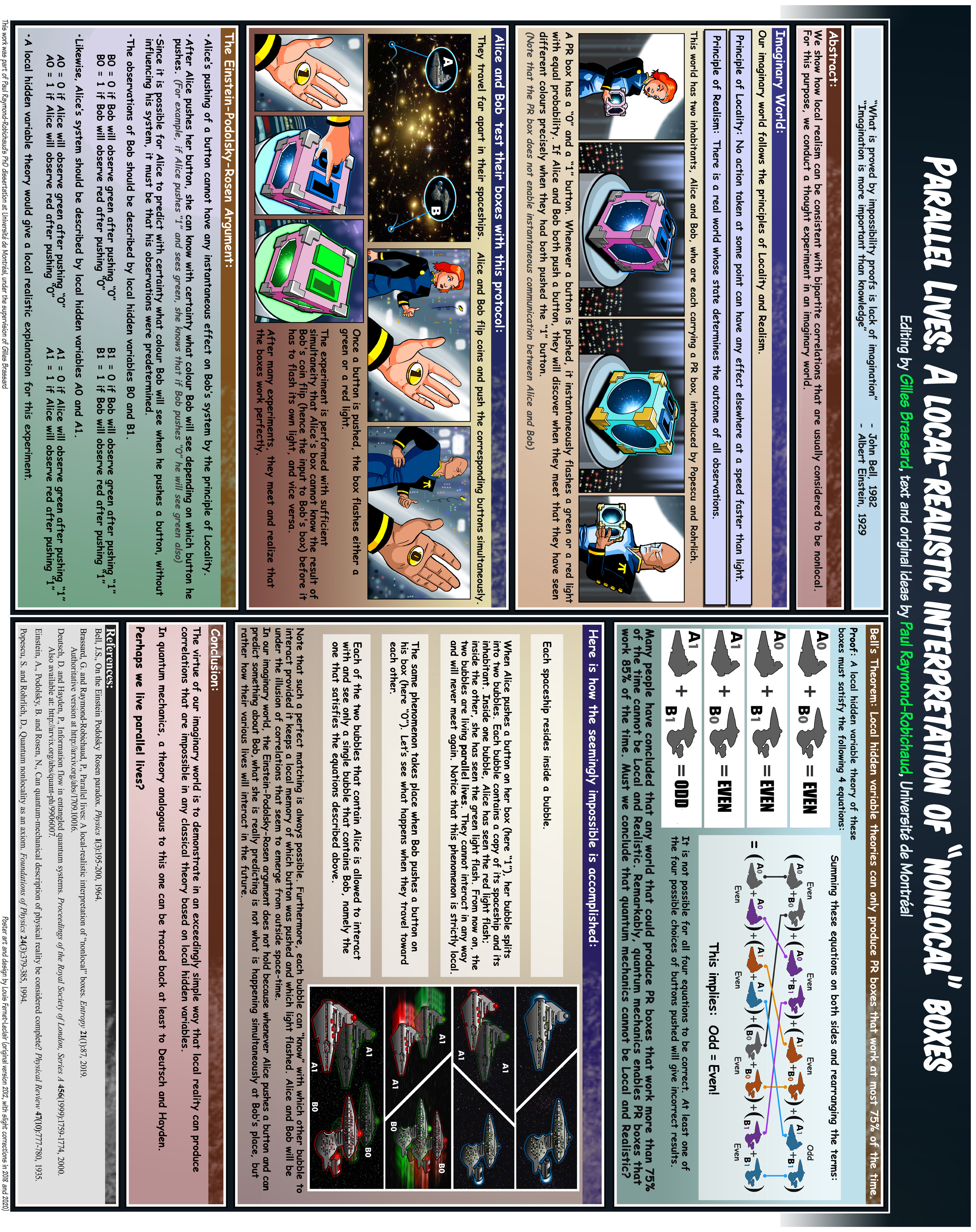}

\end{document}